# Reconfigurable all-diffractive optical filters using phase-only spatial light modulators

**Gladys Mínguez-Vega,[1, 2*] V. R. Supradeepa,[3] Omel Mendoza-Yero,[1, 2] and Andrew M. Weiner[3]**

[1] *GROC•UJI, Departament de Física, Universitat Jaume I, 12080 Castelló, Spain.*

[2] *Institut de Noves Tecnologíes de la Imatge (INIT), Universitat Jaume I, 12080 Castelló, Spain.*

[3] *School of Electrical and Computer Engineering, Purdue University, West Lafayette, Indiana 47907, USA.*

[*] *Corresponding author: gminguez@uji.es*

We demonstrate a reconfigurable optical filter implemented using a phase-only two-dimensional liquid crystal on silicon spatial light modulator. To achieve this we utilize two different approaches leading to two different configurations in the modulator. The first one, based on a spatially patterned diffractive lens permits to obtain the desired spectrum along the optical axis and, in the second one, based on a generalized spectrometer, it is found outside of the optical axis. Experimental results show good agreement with the theory and indicate the validity of this technique.

Optical filters based on diffractive optical elements (DOEs) have received increased attention since the development of the first synthetic spectrum as a tool for correlation spectroscopy [1]. The generation of a synthetic spectrum requires the design of a DOE that transforms the spectrum of the incident light into the spectrum of interest. Based on this idea, several approaches have been reported in the literature. DOE-based spectral synthesis has been performed by means of generalized spectrometers [2], digital mirror arrays [3], micromechanical grating arrays [4], micromechanical diffractive lenses [5], moving micro-gratings [6], or spatially patterned kinoform diffractive lenses (DL) [7]. In general these systems are complex, requiring the use of diffractive, refractive and in some cases expensive high technology components, that maybe difficult to align. The spatial light modulator (SLM) could provide an opportunity to change such a situation, because it allows encoding the suitable spatial patterns to generate the desired spectrum in a compact and inexpensive manner. In this Letter we demonstrate that a single phase-only two dimensional (2D) SLM can be used as an optical filter. To increase the flexibility, in the SLM we implemented two different configurations. The first one is based on spatially patterned DL [7] and the second on an all-diffractive generalized spectrometer [2, 8].

As a first optical filter we consider the system based on a spatially patterned DL [7]. This device has also shown its potential as a quasi-direct space-to-time pulse shaper [9]. We will first start with a simple formulation for this filter where we will assume that the DL and the mask are independent entities. For this setting we will describe a methodology to obtain the corresponding complex mask function having both amplitude and phase information. We will then describe the procedure to integrate this mask as well as the DL together in the SLM as a single phase pattern.

The optical filter based on a spatially patterned DL consists of a circularly symmetric mask and a DL facing each other. The DL can be approximated by a thin lens with a frequency-dependent focal length of value $Z(\nu) = Z_o \nu / \nu_o$, where $Z_o$ is the focal length for the frequency $\nu_o$. Up to now this optical filter has been used with fixed masks. Now with the SLM we generate user-defined spectra that can be reconfigured with a refresh rate of 60 hertz and above.

Let us consider a complex transmittance mask $t(r) = |t(r)| \exp[i\phi(r)]$, with $|t(r)|$ the amplitude and $\phi(r)$ the phase of the function. For convenience we express $t(r)$ as a function of a new variable, defined as $s = r^2 / a^2 - 0.5$ where $r$ refers to the radial coordinate and $a$ is the maximum extension of the mask function. In such a way we define $q(s) = t(r)$. We use a polychromatic plane-wave of power spectrum $S(\nu)$ to illuminate the optical filter constituted by the combination of this mask with a DL. Under this condition, the on-axis output irradiance $I_{out}(\nu)$ at the focal plane of the system, located at a distance $Z_o$ from the DL, is given by

$$I_{out}(\nu) = \left( \frac{\pi \nu a^2}{Z_o c} \right)^2 S(\nu) \left| \tilde{Q}\left( \frac{a^2 (\nu - \nu_o)}{2 Z_o c} \right) \right|^2 , \tag{1}$$

where $\tilde{Q}(u) = \int_{-\infty}^{\infty} q(s) \exp[i 2\pi u s] ds$ is the Fourier transform (FT) of $q(s)$. If $\Delta\nu \ll \nu_o$, the leading $\nu$ in Eq. (1) can be approximated by $\nu_o$. If this approximation does not hold, the influence of $\nu$ factor can be precompensated in the mask design.

Now we discuss the mask design. In order to find a finite aperture complex mask corresponding through eq. (1) to a user-defined spectrum, we use the four-step iterative algorithm described in [10]. We illustrate our procedure with an example in which we choose a $\tilde{Q}(u)$ function which corresponds to two equal amplitude Gaussian functions. In view of our system parameters, we choose a target comprising two peaks of the same high centered at 1550 nm and 1592 nm with a FWHM of about 11.5 nm. Figures 1 (a) and (b) show the profile of the mask function $q(s)$ for the amplitude and phase, respectively. After the conversion from the $s$ to the $r$ coordinate and the construction of the 2D mask, the amplitude and phase distribution for the mask function $t(r)$, can be shown in Fig. 1 (c) and (d), respectively.

The experimental setup is basically composed of a laser system, an SLM, and an optical spectrum analyzer. The laser system is a passively mode-locked Erbium fiber laser with a center wavelength of 1570 nm that generates pulses with a repetition rate of 50 MHz. The laser radiation is coupled from fiber to free space by using a standard collimator and sent through a free space polarizer. The polarizer allows us to select the correct polarization state required to match the sensitive axis of the SLM. A fiber based polarization controller is used before the collimator to ensure maximum power transmitted through the polarizer. Light then passes through a 10X beam expander, an iris, and a beam splitter (BS). The transmitted beam from the BS is used to illuminate the SLM. It has a relatively flat intensity profile and a diameter equal to the smaller dimension of the SLM. The SLM, working in reflection mode, allows a flexible beam shaping. It is a phase only liquid crystal on silicon device (HEO 1080 P, Holoeye) that provides $>2\pi$ phase modulation over our bandwidth. It was calibrated for 1570 nm using a standard polarimetric setup [11]. The beam is then reflected in the BS and sent back into a bare fiber tip of standard single mode fiber (core size of ~9 μm). Finally, the output power spectrum is monitored through a commercial optical spectrum analyzer (ANDO AQ6317).

To encode a complex-valued function (the complex mask transmittance $t(r)$ together with the DL) onto a phase-only SLM, several procedures have been proposed in the literature. We follow the one reported in [12]. In this technique a spatially modulated phase is encoded onto the SLM. For each pixel of the SLM the encoded phase function is obtained by multiplying the desired amplitude and the phase, i.e. the phase transmission function of the SLM is $T(r) = \exp[i|t(r)|(\phi(r) + \phi_{DL})]$, where $\phi_{DL} = \exp[i\pi v_o r^2 / Z_o c]$ is the phase introduced by the DL. In our case $Z_o$=4 cm for the wavelength of $\lambda_o$=1570nm, and we define the mask aperture by setting $|t(r)| = 0$ for r > 4.3 mm. The amplitude of the mask transmittance must be defined within the range [0, 1] whereas the total phase in range [−π ,π]. As the depth of the phase distribution changes with $|t(r)|$, we spatially modify the diffraction efficiency of the pattern. Mathematically this approach can be demonstrated by using a mixed Fourier-Taylor series to expand $T(r)$ and interpreting this expansion as the formation of different diffraction orders [12]. In our case each diffraction order will focus in a different plane along the axis. In the first diffraction order, localized at the focal distance of the DL, we place the output fiber to obtain the shaped spectrum. The main limitation of this technique is that it can appear a sinc amplitude modulation coming from the series expansion. Although this distortion can be precompensated [12], in our case it was negligible. Figure 2 (a) shows the central pixels of the phase transmission function encoded in the SLM following this procedure for the mask function shown in Fig 1.

The theoretical and experimental output power spectrum are shown in Fig. 3 (top) using continuous and dotted lines, respectively. The dashed line is the input power spectrum of the laser. In the theoretical spectrum we incorporated the effect of the power spectrum of the incoming pulse, which causes a reduction of the peak height for the shorter wavelengths. Fig. 3(a) corresponds to the output power spectrum obtained for the masks shown in Fig.2 (a). In Fig. 3(b) we adjust the output spectrum to be three Gaussian functions in $\tilde{Q}(u)$, for our systems parameters corresponding to equal amplitude peaks centered at 1546 nm, 1570 nm and 1594 nm with a FWHM of about 10 nm. The measured spectrum is again close to the target.

As a second configuration, we implement in the SLM a modified version of a generalized spectrometer [2,8]. A generalized spectrometer is composed of a diffractive grating (DG) attached to a spatially patterned mask, $m(x) = |m(x)| \exp[i\phi(x)]$, and a refractive lens that focus the light in a fiber spectrometer. This device has also shown its potential as a direct space-to-time pulse shaper [13]. The key development in our work with respect to previously reported literature is that we emulate the refractive lens by a suitable DL and we implement everything, the DG, the DL, and the mask all together in the SLM to create a compact system.

A key difference which arises when we use a DL instead of a refractive lens here is that both the DL and the DG are dispersive elements so it is necessary to implement the device in such a way that the dispersive action of the DL can be neglected in comparison with that given by the DG. To this end we take into account that the maximum spectral resolution of a DG spectrometer is provided by $\Delta\lambda_{DG} \propto \lambda_o p / d$ with $\lambda_o$ the center wavelength, $p$ the period of the DG and $d$ the grating size. In the case of a DL spectrometer the maximum spectral resolution is provided by $\Delta\lambda_{DL} \propto \lambda_o^2 2Z_o / a^2$ where $a$ is the maximum radial extension of the DL. This means that we require a large focal distance for the DL and small period for the DG to have $\Delta\lambda_{DL} >> \Delta\lambda_{DG}$. Using normal incidence to the grating, the output power spectrum located at the position of the first diffraction order of the DG for the mean frequency $\nu_o$ is then

$$I_{out}(\nu) \propto S(\nu) \left| \tilde{M}\left( \frac{(\nu - \nu_o)}{p \nu_o} \right) \right|^2 , \tag{2}$$

when plane-wave illumination is considered. In Eq. (2) $\tilde{M}$ is the FT of the mask function $m(x)$.

To encode in the phase-only SLM the complex-valued function constituted by the 1D mask, the 2D DL and the 1D DG, we follow again the procedure employed in [12]. Now, the phase transmission function of the SLM is provided by $T = \exp\left[i\left|m(x)\right|(\phi(x)+\phi_{DL}+\phi_{DG})\right]$, where $\phi_{DG} = \exp\left[i2\pi x/p\right]$. As an example, Fig. 2(b) shows the central pixels of the phase transmission function encoded in the SLM to obtain the same target as with the mask shown Fig. 2(a). Now we consider experimental parameters of $Z_o$=25 cm for the wavelength of $\lambda_o$=1570nm and $p$=32 µm. This guarantees that $\Delta\lambda_{DL} >> \Delta\lambda_{DG}$. Let us remark that with this configuration the output point is located in the first diffraction order of the DG; so it is shifted off-axis by the distance $x_o = \lambda f / p$ which corresponds here to about 1.2 cm.

For the same targets as above, the desired output power spectra, the corresponding experimental data, and the input power spectrum of the laser are shown in Fig. 3 (c) and (d) with continuous, dotted and dashed lines, respectively. We show a good agreement between the expected patterns and the measured ones. We believe that the peak height discrepancies in Fig. 3(d) are mainly due to a slight misalignment of the output fiber position.

In conclusion, we have designed and verified experimentally a compact, versatile and real time tunable optical filter using a phase only SLM. We have performed the same experiments with two different configurations in the SLM. The configuration based on the spatially patterned DL allows to work along the optical axis and is more compact. The setup based on the generalized spectrometer works off-axis and has higher spectral resolution. In both setups it would be possible to improve performance by using an active feedback that adjusts the grey levels applied on each pixel, in order to match the target filter. This research promises to have impact in different fields whenever a user defined spectrum is required.

This research was funded in part by the Spanish Ministerio de Educación y Ciencia (MEC), Spain, through Consolider Programme SAUUL CSD2007-00013. G. Mínguez-Vega acknowledges financial support to the Generalitat Valenciana for the grant BEST/2009/135.

REFERENCES

1. M. B. Sinclair, M. A. Butler, A. J. Ricco, and S. D. Senturia, “Synthetic spectra: a tool for correlation spectroscopy,” Appl. Opt. **36**, 3342-3348 (1997).
2. J. D. McKinney and A.M. Weiner, “Engineering of the passband function of a generalized spectrometer,” Opt. Express **12**, 5022-5036 (2004).
3. M. Kanpczyk, A. Krishnan, L. Grave de Peralta, A. A. Bernussi, and Temkin, “Reconfigurable optical filter based on digital mirror arrays,” IEEE Photon. Technol. Lett. **17**, 1743-1745 (2005).
4. H. Sagberg, M. Lacolle, I.R. Johansen, O. L∅vhaugen, R. Belikov, O. Solgaard, and A. S. Sudb∅, “Micromechanical gratings for visible and near-infrared spectroscopy,” IEEE J. Selec. Topics Quantum Electron. **10**, 604-613 (2004).
5. M. Lacolle, H. Sagberg, I.R. Johansen, O. L∅vhaugen, O. Solgaard, A. S. Sudb∅, “Reconfigurable near-infrared optical filter with a micromechanical diffractive lens,” IEEE Photon. Technol. Lett. **17**, 2622-2624 (2005).
6. G. Zhou and F. S. Chau, “Spectral synthesis using an array of micro gratings,” Opt. Express **16**, 9132-9143 (2008).
7. G. Mínguez-Vega, O. Mendoza-Yero, E. Tajahuerce, J. Lancis, and P. Andrés, “Optical filter based on a spatially patterned kinoform diffractive lens,” IEEE Photon. Technol. Lett. **21**, 347-349 (2009).
8. D. Leaird and A. M. Weiner, IEEE J. of Quantum Electron. **37**, 494-504 (2001)
9. G. Mínguez-Vega, O. Mendoza-Yero, J. Lancis, R. Gisbert, and P. Andrés, “Diffractive optics for quasi-direct space-to-time pulse shaping,” Opt. Express **16**, 16993-16998 (2008).

10. J. A. Davis, C.S. Tuvey, O. López-coronado, J. Campos, M. J. Yzuel, and C. Iemmi, "Tailoring the depth of focus for optical imaging systems using a Fourier transform approach," Opt. Lett. **32**, 844-846 (2007).

11. A.M. Weiner, *Ultrafast Optics,* p. 371, (Wiley, 2009).

12. J. A. Davis, D. M. Cottrell, J. Campos, M. J. Yzuel, and I. Moreno, "Encoding amplitude information onto phase-only filters," Appl. Opt. **38**, 5004-5013 (1999).

13. D. E. Leaird and A. M. Weiner, "Femtosecond optical packet generation by a direct space-to-time pulseshaper," Opt. Lett. **24**, 853-855 (1999).

FIGURE CAPTIONS

1. Profile of the mask function in the normalized coordinate *s* for: (a) the amplitude and (b) phase distribution. 2D circular symmetric mask in the radial coordinate for: (c) amplitude and (d) phase distribution.
2. Phase transmission function encoded in the central 500 x 500 pixels of the SLM for: (a) The spatially patterned DL and (b) the generalized spectrometer based optical filters.
3. Measured (dotted line) and theoretical (continuous line) power spectra obtained (a,b) with the spatially patterned DL optical filter [(a) two peak spectrum, (b) three peak spectrum] and (c,d) with the generalized spectrometer optical filter [(c) two peak spectrum, (d) three peak spectrum]. The dashed line corresponds to the input power spectrum.

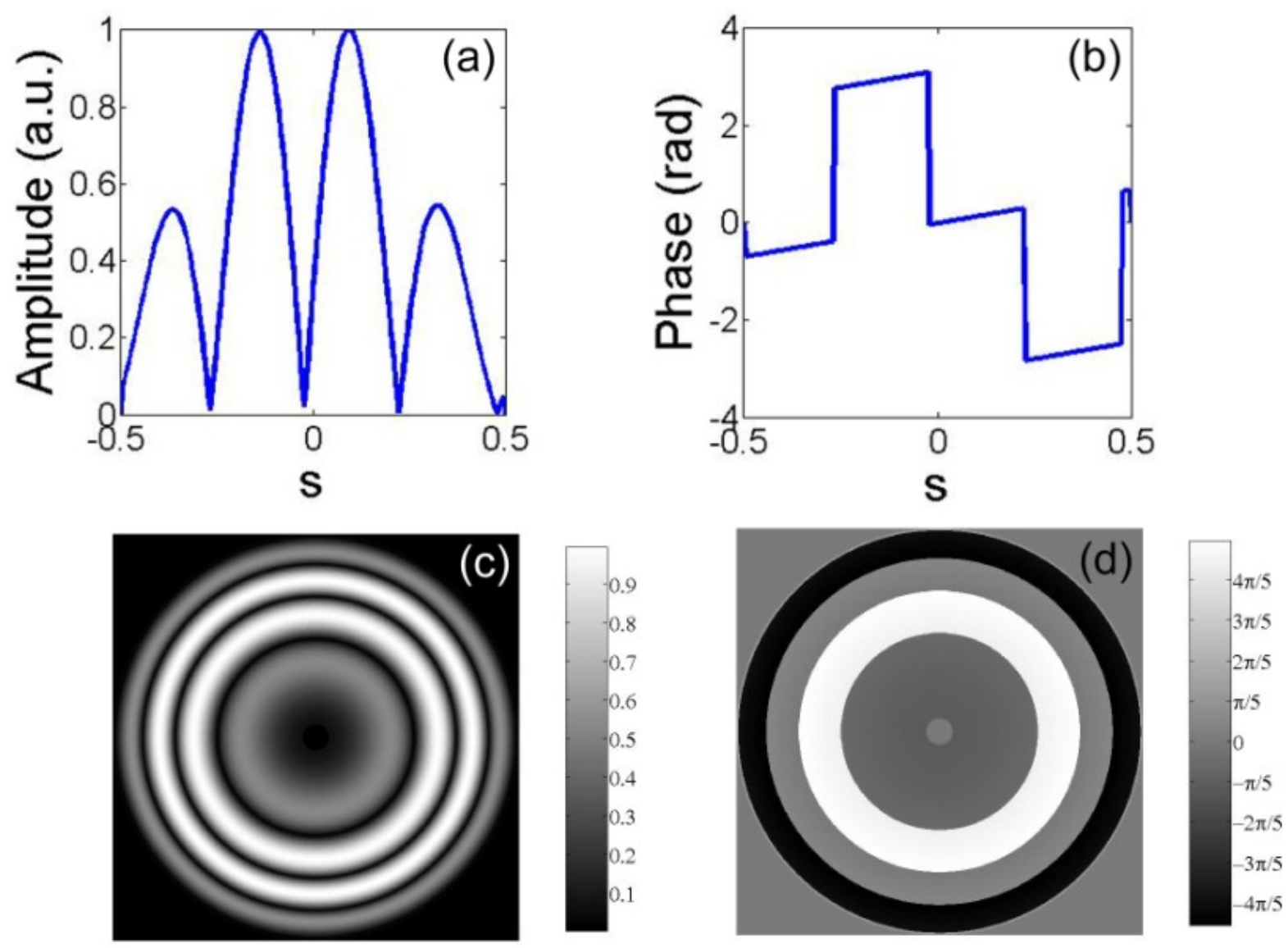

Fig. 1. Profile of the mask function in the normalized coordinate *s* for: (a) the amplitude and (b) phase distribution. 2D circular symmetric mask in the radial coordinate for: (c) the amplitude and (d) phase distribution.

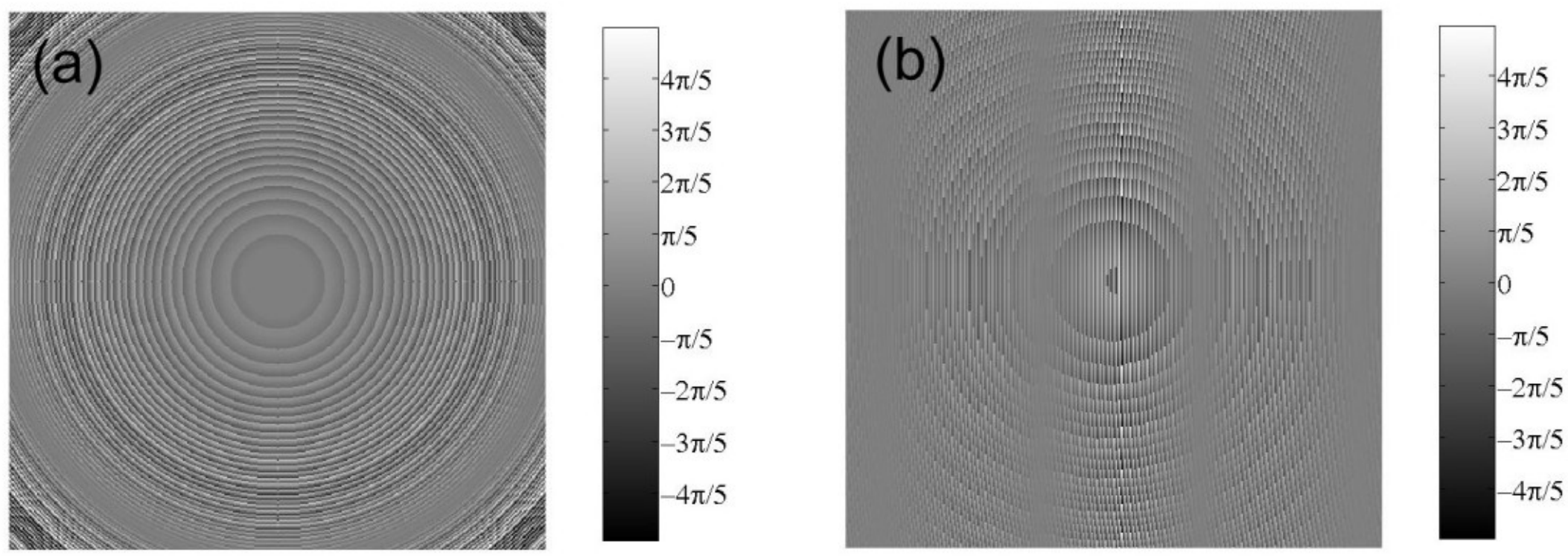

Fig. 2. Phase transmission function encoded in the central 500 x 500 pixels of the SLM for: (a) The spatially patterned DL and (b) the generalized spectrometer based optical filters.

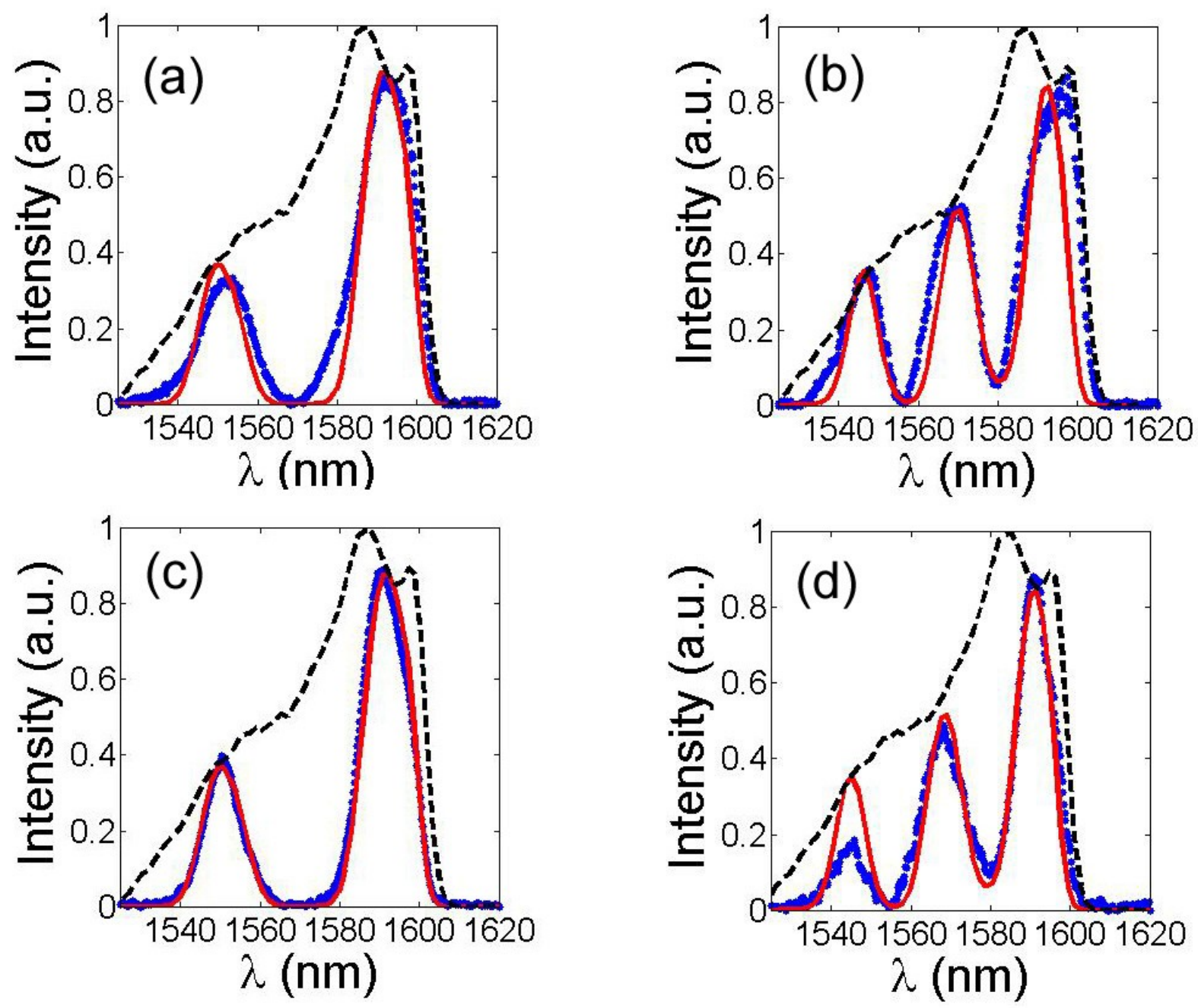

Fig. 3. Measured (dotted line) and theoretical (continuous line) power spectra obtained (a,b) with the spatially patterned DL optical filter [(a) two peak spectrum, (b) three peak spectrum] and (c,d) with the generalized spectrometer optical filter [(c) two peak spectrum, (d) three peak spectrum]. The dashed line corresponds to the input power spectrum.